\documentclass[twocolumn,aps,pra,longbibliography,superscriptaddress]{revtex4-2}
\usepackage{amsmath,latexsym}
\usepackage{xcolor}
\usepackage[%
  colorlinks=true,
  urlcolor=blue,
  linkcolor=blue,
  citecolor=blue
]{hyperref}
\usepackage{etoolbox}
\usepackage{breqn}
\usepackage{mathrsfs,verbatim,mathtools,graphicx,amsmath, amssymb, bm, epstopdf, enumerate}
\makeatletter
\let\cat@comma@active\@empty
\makeatother

\begin{document}

\preprint{APS/123-QED}

\title{Fast control of the transverse structure of a light beam using acousto-optic modulators}

\author{Mahdieh Chartab Jabbari}
\email{m.jabbari1993@gmail.com}
\affiliation{Department of Physics, University of Ottawa, Ottawa, Ontario K1N 6N5, Canada}
\author{Cheng Li}
\affiliation{Department of Physics, University of Ottawa, Ottawa, Ontario K1N 6N5, Canada}
\author{Xialin Liu}
\affiliation{Department of Physics, University of Ottawa, Ottawa, Ontario K1N 6N5, Canada}
\author{R. Margoth Córdova-Castro}
\affiliation{Department of Physics, University of Ottawa, Ottawa, Ontario K1N 6N5, Canada}
\author{Boris Braverman}
\affiliation{Department of Physics, University of Toronto, Toronto, ON, M5S1A7, Canada}
\author{Jeremy Upham}
\affiliation{Department of Physics, University of Ottawa, Ottawa, Ontario K1N 6N5, Canada}
\author{Robert W. Boyd}
\email{rboyd@uottawa.ca}
\affiliation{Department of Physics, University of Ottawa, Ottawa, Ontario K1N 6N5, Canada}
\affiliation{The Institute of Optics, University of Rochester, Rochester, New York 14627, USA}

\date{\today}
\begin{abstract}
Fast, reprogrammable control over the transverse structure of light beams plays an essential role in applications such as structured illumination microscopy, optical trapping, and quantum information processing. Existing technologies, such as liquid crystal on silicon spatial light modulators (LCoS SLMs) or digital micromirror devices (DMDs), suffer from limited refresh rates, low damage thresholds, and high insertion loss. Acousto-optic modulators (AOMs) address these limitations and enable rapid phase and amplitude modulation controlled by the amplitude and frequency of the RF waveform. By effectively mapping the temporal RF waveforms onto the spatial diffraction patterns of the optical field, individual AOMs have been shown to generate one-dimensional (1D) spatial modes at a pixel refresh rate of nearly 20 MHz \cite{liu2023using}. Here, we extend this concept to enable fast modulation in a two-dimensional (2D) space using a double-AOM scheme and show the generation of 2D Hermite-Gaussian (HG$_{nm}$) modes. Numerical simulations demonstrate that one can improve the fidelity to a desired amplitude and phase pattern for the diffracted beam by optimizing the amplitude and frequency of the input RF waveform. With optimized RF input, our experimentally generated HG modes have an average mode fidelity of 81\%, while the highest order mode generated, HG$_{53}$, retains a fidelity of 56\%.
\end{abstract}

\maketitle


\section{\label{sec:level1}Introduction}

The generation and manipulation of spatial light modes play essential roles in a wide range of applications. Access to multiple spatial modes allows encoding of information in a higher dimension, which enhances the capacity of quantum information processing \cite{kagalwala2017single,brandt2020high, xavier2020quantum}. The ability to shape optical wavefronts helps to create complex optical traps, enabling novel structures of optical tweezers and laser material processing technologies \cite{yang2021optical,mohl2019tailored}.

Diffractive optical elements, such as LCoS-SLMs, and mechanical optical components, such as DMDs, are typically used to generate and manipulate the spatial structure of light. LCoS-SLMs offer high resolution and programmable modulation but are usually limited to a refresh rate of around 120 Hz due to the long response time of liquid crystal molecules. Using liquid crystals also results in low damage thresholds of the order of microjoules per square centimeter \cite{xing2020high, turtaev2017comparison, braverman2020fast}. In contrast, DMDs, which consist of an array of microelectromechanical mirrors, can achieve refresh rates of up to 22 kHz \cite{liu2023using, mitchell2016high} but exhibit greater insertion loss than LCoS-SLMs. 

Acousto-optic modulators (AOMs), consisting of crystals driven by piezoelectric transducers, can potentially overcome the limitations mentioned above \cite{chizhikov2022high}. LCoS-SLMs update the generated spatial mode in a time interval on the order of several milliseconds \cite{zhou2020laser, braverman2020fast,turtaev2017comparison}, while AOMs can modulate at frequencies up to tens of MHz, a bandwidth associated with single-pixel modulation. The effective refresh rate for full spatial patterns is set by the acoustic transit time across the beam waist. This fast refresh rate makes AOMs suitable for applications that require high-speed modulation, such as laser scanning microscopy and optical communications \cite{duocastella2020acousto, pohl2013advances}. Furthermore, typical materials comprising AOMs, such as fused silica (SiO$_2$) and tellurium dioxide (TeO$_2$), offer optical damage thresholds of the order of joules per square centimeter (J/cm$^2$), which are several orders of magnitude higher than those of LCoS-SLMs. This advantage makes AOMs more suitable for handling high-power laser beams. The diffraction efficiency of AOMs typically ranges from 50\% to 90\% or higher, exceeding that of DMDs \cite{zhang2019beam,nikitin2022acousto}. Unlike LCoS-SLMs, which are sensitive to the polarization of incident laser beams, the diffraction efficiency of AOMs is unaffected by polarization \cite{lazarev2012lcos,igasaki1999high}. 

Spatial light modulation using AOMs has found applications in various fields. Treptow et al. eliminated common coherent artifacts in holography by harnessing the intrinsic motion of acousto-optic holograms \cite{treptow2021artifact}. A double-pass AOM, functioning as a fast steerable mirror with a high switching rate, facilitates rapid state tomography of spatial modes within a two-dimensional (2D) Hilbert space \cite{braverman2020fast}. Recent developments in acousto-optic systems have also shown potential for random access scanning, ultrafast confocal and multiphoton imaging, as well as fast inertia-free light-sheet microscopy \cite{salome2006ultrafast, duocastella2017fast, duocastella2020acousto}. Despite their advantages, AOMs have been used mainly in one-dimensional (1D) spatial modulation. To access the full potential of AOM-based spatial light manipulation, a scheme for independent control over orthogonal components of the optical field in a 2D space is essential.


This paper proposes and demonstrates a double-pass AOM scheme that modulates an optical beam in orthogonal transverse directions. We first numerically simulate the optimal RF waveform parameters for generating high-fidelity 2D Hermite-Gaussian (HG$_{nm}$) modes. Using the optimized parameters, we experimentally demonstrate the generation of HG$_{nm}$ modes up to a total mode order of n + m = 8. By optimizing the input RF waveform, the fidelity of the generated modes to the target modes reaches an average of 81\%. Although the fidelity of the generated mode relative to the target decreases with increasing mode order, the experimentally generated modes retain a fidelity of 56\% for the highest-order target mode (HG$_{53}$). 

\section{\label{sec:level1}Principles of operation}

\begin{figure}[t]
	\centering\includegraphics[scale=0.38]{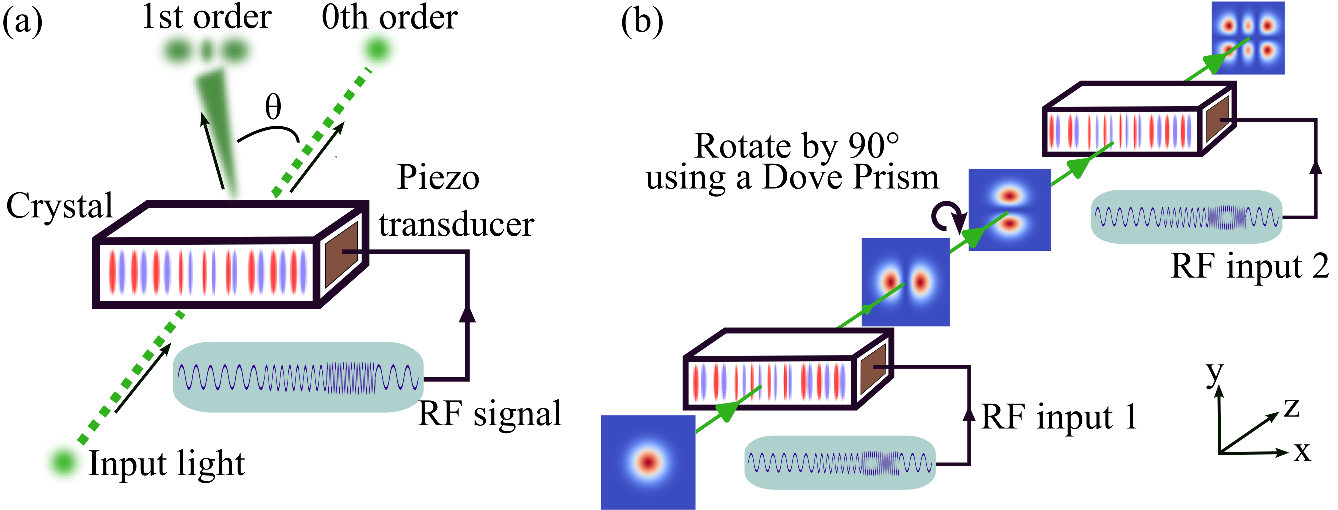}
	\caption{(a) A schematic diagram of AOM-based spatial light modulation. Blue and red false-color blobs show the crystal being compressed or stretched by the acoustic wave. (b) The principle of generating 2D structured patterns from two 1D modulations.}
    \label{fig1}
\end{figure}

Fig.~\ref{fig1}(a) depicts the schematic of an AOM in which a piezoelectric transducer converts an RF waveform into a sound wave propagating through a TeO$_2$ crystal. The compression pattern induced by the sound wave creates a spatially varying refractive index, as illustrated by a false color pattern. This refractive index pattern acts as a diffraction grating that imprints a spatial structure onto the diffracted beam. Adjusting the amplitude and frequency of the RF waveform thus manipulates the amplitude and phase of this beam. AOMs typically accommodate RF drive frequencies between 50 and 200 MHz \cite{liu2023using}. 


The first-order diffraction, which exhibits the highest diffraction efficiency, is generated at an angle $\theta$ given by 

\begin{equation}
    \sin\theta= \frac{{\lambda_0}}{{n \lambda_s}}= \frac{{\lambda_0 f}}{{n v_s}},
\end{equation}
where $n$ represents the refractive index of the crystal, $\lambda_0$ and $\lambda_s$ denote the wavelengths of light in vacuum and sound, respectively, $v_s$ is the velocity of sound propagation, and $f$ is the carrier frequency of the RF waveform. The input beam should enter the AOM at the Bragg angle, $\theta_B=\theta/2$, to achieve the highest diffraction efficiency. Light diffraction through an AOM obeys Bragg’s law. The Bragg condition described here refers to the internal angle between the optical beam and the acoustic wave inside the crystal, where the phase-matching condition is satisfied \cite{saleh2019fundamentals}.

The time-dependent amplitude of the RF waveform is mapped directly onto the intensity profile of the optical beam, whereas the variation of the RF frequency modulates the phase pattern. Since beam components diffracted into different angles accumulate phase differences as they propagate along different paths inside the AOM, the desired 1D modulation of the complex field amplitude at the output plane can be obtained by controlling the temporal variation of the RF frequency driving the AOM. It is important to note that this description applies at a given instant in time, making the approach particularly relevant for pulsed light, where the optical pulse effectively samples the instantaneous RF waveform.

The complex profile of the modulated beam $E(d)$ is determined by the position-dependent amplitude $A(d)$ and phase $\phi(d)$ of the light field. In the thin-AOM limit, where diffraction during light propagation through the crystal is negligible, $A(d)$ approximately replicates the amplitude of the RF waveform. Under this condition, the overlap between a modulated beam and the desired spatial mode is mainly determined by $\phi(d)$, which is related to the position-dependent frequency detuning from the Bragg frequency $\Delta f(d)$ through

\begin{equation}
\phi(d) \sim \frac{2\pi d}{v_s} \, \Delta f(d),
\end{equation}
where $d$ indicates the position in the AOM along the propagation direction of the sound wave.

To enable light modulation in 2D space, a common approach is to mount two AOMs at $90^{\circ}$ with respect to each other so that they independently modulate the beam along perpendicular directions. A common approach is to mount the two AOMs at $90^{\circ}$ with respect to each other. An alternative approach is to use a Dove prism to rotate the field between the two AOMs. Instead, we rotate the beam diffracted by the first AOM by $90^{\circ}$ using a Dove prism (DP) before sending it to the second AOM, as illustrated in Fig.~\ref{fig1}(b). We denote the transfer matrices of the first and second AOMs as $\hat{U}_{x1}$ and $\hat{U}_{x2}$, and that of DP as a rotation operator $\hat{R}$. The output field can then be expressed as 
\begin{equation}
    E_o = \hat{U}_{x2}\hat{R}\hat{U}_{x1}E_i,
\end{equation}
where $E_i$ represents the input optical field; notice that the y-direction modulation is realized by a rotated x-direction modulation, namely, $\hat{U}_y = \hat{R}\hat{U}_x$. This approach enables light modulation in both the x and y directions $E_o = \hat{U}_x\hat{U}_yE_i$, while keeping the two AOMs aligned in parallel. 

When driving AOM$_1$ and AOM$_2$ with RF waveforms $f(x_1)$ and $f(x_2)$ respectively, the output field $E_o(x,y)$ is then given by $E_o[f(x_1), f(x_2);x,y] = \hat{U}_{x2}[f(x_2)]\hat{R}\hat{U}_{x1}[f(x_1)]E_i$. We can then optimize the modulation pattern by jointly iterating \( f(x_1) \) and \( f(x_2) \) to maximize fidelity, which is defined as
\begin{widetext}
\begin{equation}
\mathcal{F}\!\left[f(x_1), f(x_2)\right]
=
\frac{\left|\iint E_t^{*}(x,y)\,
E_o\!\left[f(x_1), f(x_2);x,y\right]\, dx\, dy\right|^2}
{\iint |E_t(x,y)|^2\, dx\, dy \;
 \iint |E_o\!\left[f(x_1), f(x_2);x,y\right]|^2\, dx\, dy}
\end{equation}
\end{widetext}
\begin{equation}
\left[f(x_1), f(x_2)\right]
=
\arg\max_{f(x_1),\, f(x_2)}
\mathcal{F}\!\left[f(x_1), f(x_2)\right]
\end{equation}
where \(E_t(x,y)\) represents the target electric field.


\section{\label{sec:level1}Numerical simulations}

We numerically simulate the generation of HG modes by starting with a Gaussian input beam and computing the resulting diffraction patterns using the theoretical framework described above. To numerically model the AOM-based beam shaping, we treat the acousto-optic crystal as a volume Bragg grating whose refractive-index modulation is directly determined by the applied RF waveform. The temporal RF signal is mapped to a spatial index profile across the interaction region, and the optical field is propagated through the crystal using a split-step Fourier beam-propagation method. This approach captures the cumulative phase shift and diffraction that occur inside the AOM and follows the simulation framework presented in \cite{braverman2020fast}, where the RF waveform controls both amplitude and phase through the induced refractive-index pattern. We consider that the AOM crystal made of TeO\(_2\) has a refractive index of 2.3 at 532~nm and a sound speed of \( v_s = 3.6~\text{km/s} \), and we assume a spatial sampling step of 1~\(\mu\)m along the propagation direction of the acoustic wave. For the optical beam propagation, we use a longitudinal step size of 5~\(\mu\)m. We also assumed an AOM aperture of \(7~\text{mm} \times 2~\text{mm}\), consistent with the effective beam size.

\begin{figure}[t]
        \centering\includegraphics[scale=0.41]{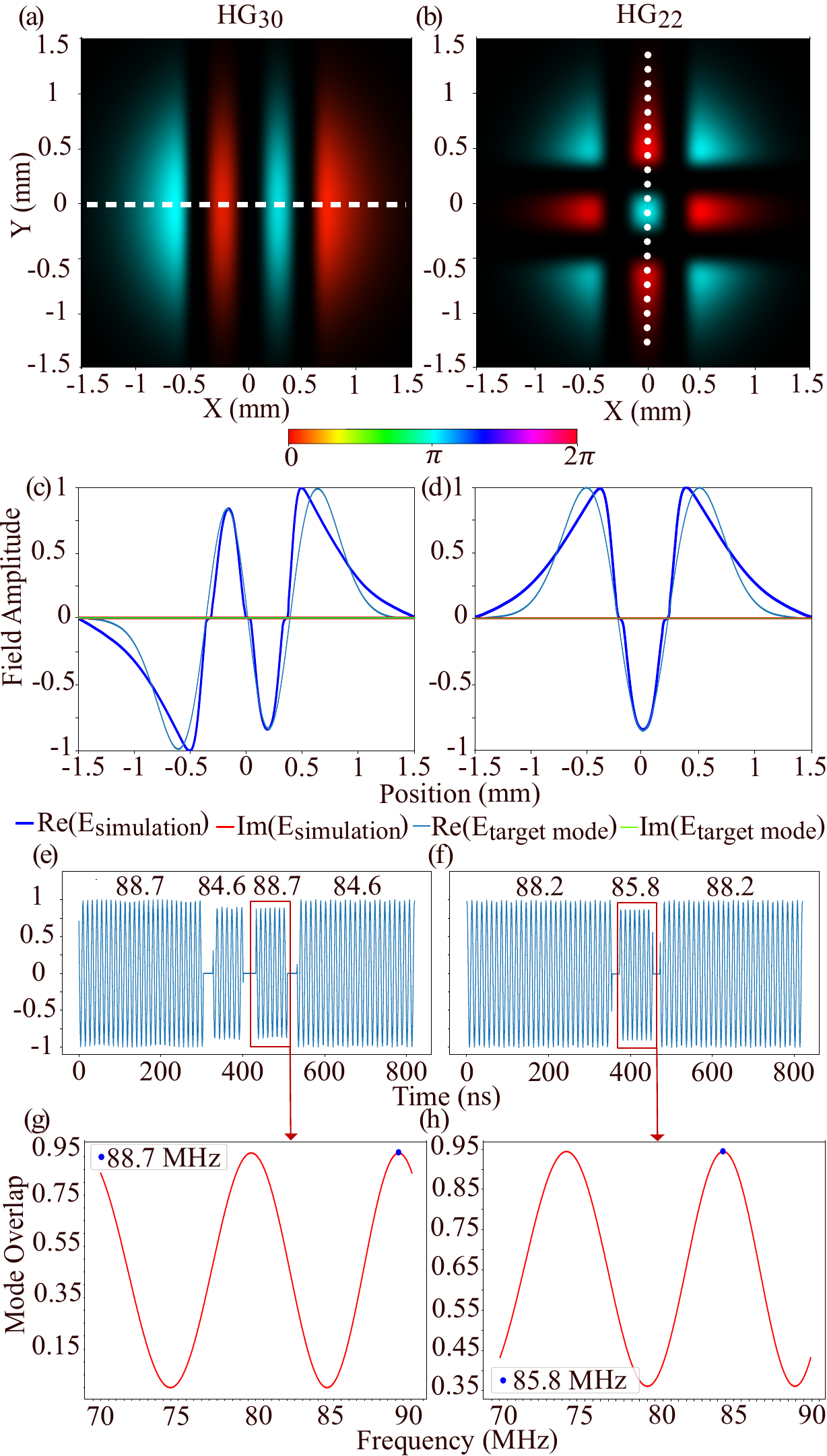}
	\caption{(a, b) Simulated intensity and phase distributions of HG$_{30}$ and HG$_{22}$ beams generated using two AOMs. (c, d) Real and imaginary components of the electric field along the horizontal dashed line in (a) and the vertical dotted line in (b), respectively, compared with the corresponding target HG modes. (e, f) RF waveforms applied to AOM2 for HG$_{30}$ and to AOM1 for HG$_{22}$. (g, h) Mode overlap versus frequency, showing optimization of the third segment of RF2 for HG$_{30}$ and the second segment of RF1 for HG$_{22}$.}
    \label{fig2}
\end{figure}

Fig.~\ref{fig2} illustrates the generation of HG$_{30}$ and HG$_{22}$ modes using two perpendicular AOMs that independently modulate the beam along the $x$- and $y$-directions. Fig.~\ref{fig2}(a) and 2(b) show the simulated intensity and phase profiles of HG$_{30}$ and HG$_{22}$, respectively. The electric fields are computed in the image plane, corresponding to the plane reached after the field propagates through the AOM crystal and then through the downstream free-space imaging section. Fig.~\ref{fig2}(c) and (d) present the real and imaginary parts of the electric field along the horizontal white dashed line in Fig.~\ref{fig2}(a) and the vertical white dotted line in Fig.~\ref{fig2}(b), respectively, compared with theory. Fig.~\ref{fig2}(e) and (f) display the RF waveforms applied to AOM$_2$ for $x$-direction modulation and to AOM$_1$ for $y$-direction modulation. Each waveform consists of multiple segments, with each segment corresponding to one lobe of the output mode. The RF frequency of each segment determines the phase of the corresponding optical lobe, with the local phase shift proportional to the frequency deviation from the Bragg frequency [Eq.~(2)], while the RF amplitude controls the lobe intensity. Finite gaps are introduced between RF segments to suppress phase distortion near the transition regions. When segments are placed too closely, the acoustic profile introduces unintended phase gradients within each lobe, reducing phase uniformity and mode fidelity. The gap length was empirically optimized to maximize agreement between the generated and target modes. Frequency optimization was then performed for each segment individually while keeping the others fixed, with the search centered on the Bragg frequency of the AOM (80~MHz), where diffraction efficiency and phase control are optimal under experimental conditions. Fig.~\ref{fig2}(g) and (h) illustrate this optimization for the third segment of RF$_2$ and the second segment of RF$_1$, respectively. This approach enables precise two-dimensional control over phase and amplitude, allowing high-fidelity generation of structured beams.

\begin{figure}[t]
	\centering\includegraphics[scale=0.18]{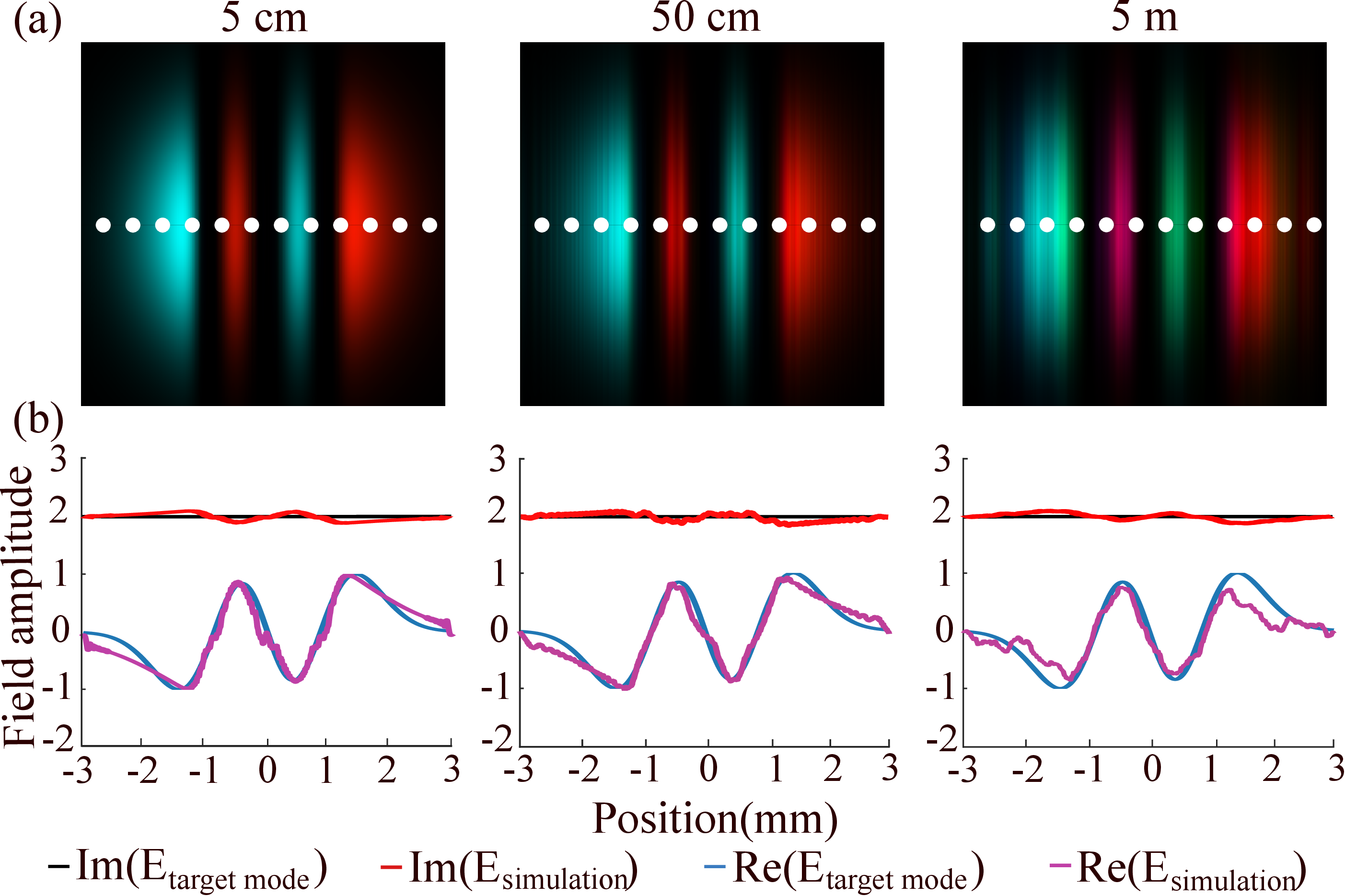}
	\caption{(a) Simulated HG$_{30}$ mode at propagation distances of 5~cm, 50~cm, and 5~m beyond the image-plane. Saturation and hue represent the amplitude and phase of the field, respectively. (b) Cross-sections of the simulated and theoretical profiles of the real and imaginary components of the electric field for each plane, with the positions of the cuts indicated by white dotted lines in (a). The imaginary components are offset vertically by two units to enhance visualization.}
    \label{fig3}
\end{figure}

We assessed the stability of the generated modes by simulating the free-space propagation of HG$_{30}$ from 5~cm to 5~m beyond the image plane. As shown in Fig.~\ref{fig3}(a), the mode structure and characteristic $\pi$-phase shifts are preserved at short distances (5~cm and 50~cm), while at 5~m the beam begins to approach one Rayleigh length ($z_R \approx 5.9$~m for a 2~mm input beam diameter), marking the onset of diffraction broadening and partial lobe overlap. Cross-sections in Fig.~\ref{fig3}(b) confirm good agreement with theory in the near field and illustrate the distortions that accumulate with propagation. Although free-space propagation is unitary and thus mode fidelity remains unchanged, the visible profile degrades due to diffraction and small $k$-vector differences introduced by the AOM grating.

\section{\label{sec:level1}Experimental Results}

\begin{figure*}[t]
	\centering\includegraphics[scale=1.2]{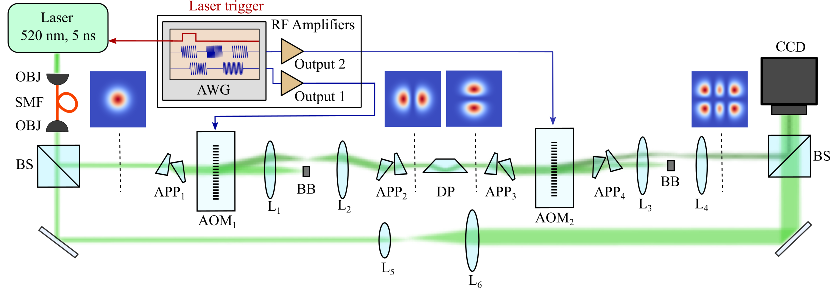}
	\caption{Schematic of the experimental setup. SMF: single-mode fiber, BS: beam splitter, APP: anamorphic prism pair, BB: beam block, DP: Dove prism. The insets show the mode profiles being modified towards the target mode, HG$_{21}$.}
    \label{fig4}
\end{figure*}

Fig.~\ref{fig4} shows the schematic of the experimental setup. A laser (Thorlabs NPL52B) emits light pulses with a duration of 10 ns at a wavelength of 520 nm and a repetition rate determined by external triggering up to 10 MHz. Light pulses are first coupled into a single-mode fiber to ensure a Gaussian-shaped spatial profile. Subsequently, the light is coupled out into free space and split into two paths, a signal and a reference. 

Since both AOMs have rectangular apertures of 7 mm by 2 mm, the signal beam is expanded to an elliptical shape using an anamorphic prism pair (APP$_1$) to fit within AOM$_1$. The elliptical beam, diffracted and modulated by AOM$_1$, is first restored to a circular shape by APP$_2$ and then rotated $90^{\circ}$ using a Dove prism, shaped by APP$_3$ into elliptical and sent to AOM$_2$. The modulated beam is then restored to a circular shape by APP$_4$. A 4-f imaging system, consisting of lenses L$_1$ and L$_2$ with focal lengths of 100 mm, places AOM$_2$ in the near field of AOM$_1$. The AOMs are driven independently by RF signals sent from different output ports of an arbitrary waveform generator (AWG, Keysight M8195A). Opaque beam blocks (BB) block the zeroth-order diffraction from each AOM. The first-order diffracted light is collimated using lenses L$_3$ and L$_4$, with focal lengths of 50 mm and 75 mm, respectively, and then directed to the rest of the setup. Lenses L$_5$ and L$_6$, with focal lengths of 15~mm and 125~mm, respectively, expand the reference beam in a Galilean telescope configuration. A beam splitter then recombines the expanded beam with the structured signal beam. A charge-coupled device camera captures the final beam profile after propagation through the entire modulation and imaging system. The complex amplitude of the signal beam is then reconstructed using digital off-axis holography \cite{massig2002digital}. The process in off-axis holography begins by applying a Fourier transform to the interference image to move into spatial frequency space; next, we filter the spatial frequency components to retain only the first-order sideband, which contains all the information about the complex field of the signal. An inverse Fourier transform of the filtered component reconstructs the amplitude and phase characteristics of the signal. 


\begin{figure}[h]
	\centering\includegraphics[scale=0.15]{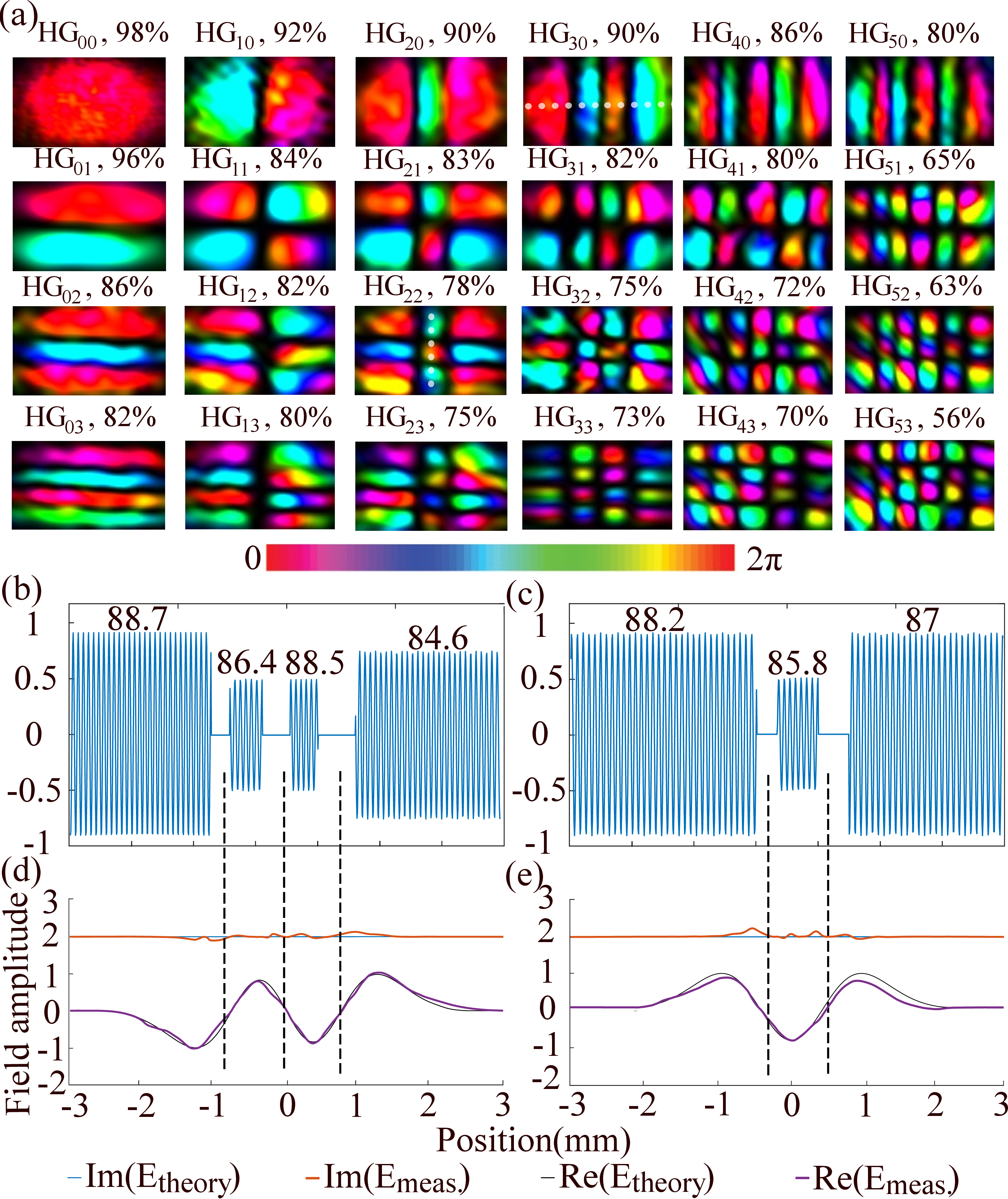}
	\caption{(a) Experimentally generated HG modes. Saturation and hue represent the intensity and phase of the field, respectively. (b) RF waveform $f(x_2)$ loaded onto $\text{AOM}_2$ to modulate the x-component of HG$_{30}$ and (c) $f(x_1)$ loaded on  $\text{AOM}_1$ to modulate the y-component of HG$_{22}$. Numbers indicate the corresponding RF frequency (MHz) in each waveform segment. Cross-sections of the measured and theoretical profiles of the field amplitude at the output of the AOM (d) for $\text{HG}_{30}$ and (e) for $\text{HG}_{22}$, with the positions of the cuts in the experimental data shown by white dotted lines in (a). The imaginary components are offset vertically by two units to enhance visualization.}
    \label{fig5}
\end{figure}

Fig.~\ref{fig5} illustrates the generated 2D HG$_{nm}$ modes for $n = 0$ to $5$ and $m = 0$ to $3$, up to the highest total mode order of $N$ = $n$ + $m$ = $8$. At very high modulation frequencies, the refractive-index pattern within the AOM may not accurately follow the applied RF waveform. In the thick-crystal regime, the departure from the ideal thin-grating approximation further contributes to phase distortions in the diffracted beam. These distortions can be mitigated by using RF waveforms with slowly varying amplitude and frequency segments, ensuring gradual spatial phase transitions within the diffracted beam. However, generating arbitrary transverse patterns often involves discontinuous phase changes, such as $\pi$ phase changes in HG modes, which can introduce significant deviations between the output field ($E_o$) and the target field ($E_t$), depending on the waveform structure and phase transitions.

We performed an experimental optimization process to maximize the mode overlap for each mode. This process follows principles similar to the numerical approach. Numerical simulations provide general guidance, but do not account for experimental imperfections, requiring fine-tuning through experimental adjustments. 

Fig.~\ref{fig5}(b) shows the optimized RF waveform $f(x_2)$ loaded on AOM$_2$ to modulate the input beam in the x direction and generate the HG$_{30}$ mode, while AOM$_1$ was driven by a constant waveform for modulation in the y direction. Each of the four distinct sections of an HG$_{30}$ mode corresponds to a segment of the RF waveform with a different frequency. Each segment is optimized individually, one after the other. We scaled the RF amplitude and adjusted the width of each segment, without further optimization, to correct for the variation in the beam amplitude between lobes. This variation is caused by the frequency-dependent diffraction efficiency, which is highest at the Bragg frequency. Fig.~\ref{fig5}(d) illustrates the real and imaginary parts of the optimized mode and a comparison with the target mode profiles, yielding a mode overlap of 90\%. The white dotted lines in Fig.~\ref{fig5}(a) indicate the cross section where the mode profiles are plotted. For 2D HG modes, the same procedure is followed for modulation in the y direction after optimizing the x direction. Fig.~\ref{fig5}(c) shows the optimized RF waveform $f(x_1)$, loaded onto AOM$_1$ to generate the HG$_{22}$ mode through $y$-direction modulation, which achieves a measured 2D overlap of 78\%.

In our setup, the acoustic displacement during a single 10~ns laser pulse is only about 0.04~mm, which is much smaller than the spatial features of the generated modes, ensuring that the acoustic pattern remains effectively unchanged during each pulse. This implies that the acoustic pattern is very close to invariant during each pulse; thus, the spatial modulation is stable and clean. While the spatial pattern of the phase is determined by the traveling acoustic wave and therefore still depends on the temporal phase of the RF waveform, breaking the RF frequencies into piecewise-constant frequency intervals reduces sensitivity to the phase of the RF frequency. In this case, small timing fluctuations simply induce a small offset in the global phase, rather than significantly distorting the phase map. Thus, while timing alignment is necessary, the requirements are less stringent compared to phase modulation schemes.

\begin{figure}[t]
	\centering\includegraphics[scale=0.22]{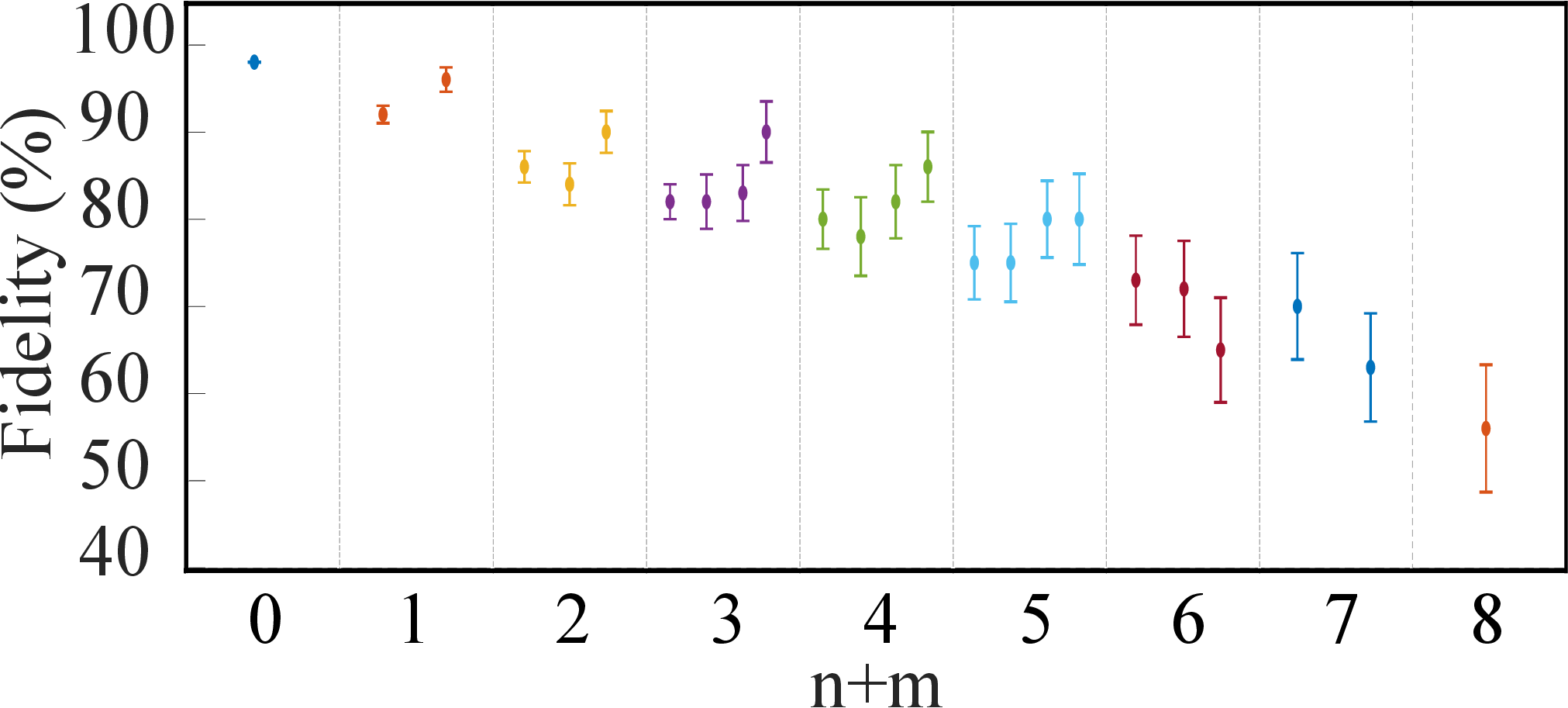}
	\caption{Dependence of the mode overlap of the generated mode on the sum $n$ + $m$ of the mode indices. Points within each $n+m$ group are ordered left to right by increasing $n$ (e.g., $n+m=3$: HG$_{30}$, HG$_{21}$, HG$_{12}$, HG$_{03}$).}
    \label{fig6}
\end{figure}

Fig.~\ref{fig6} illustrates the mode overlap in creating the desired mode for the total mode order $n+m$. We estimate the uncertainty in mode overlap using a Monte-Carlo method, where Poissonian noise is added to the acquired images to simulate experimental variations, and the standard deviation of the resulting overlap values is taken as the error \cite{mooney1997montecarlo}. Higher-order modes exhibit increased sensitivity to Poisson noise, as finer features and lower local intensity enhance relative fluctuations and overlap uncertainty. The highest-order mode studied, HG$_{53}$, retains a mode overlap of 56\%. We attribute the overall decrease in mode overlap to two primary factors. First, the diffraction efficiency can saturate and distort the intensity profile or decrease when the RF frequencies deviate from the Bragg frequency, particularly for higher-order modes requiring multiple RF frequency components. We measure the diffraction efficiencies of AOM$_1$ and AOM$_2$ to be 90\% and 82\%, respectively, at the Bragg RF frequency. As a result, the diffraction efficiency of HG$_{00}$ is 73\%; while at the highest order mode generated, HG$_{53}$, the total diffraction efficiency drops to 13\%. The second factor is the limited spatial resolution, which is determined by the total number of acoustic wavelengths that fit within the AOM aperture. This quantity is given by a product of $\delta t$, the time duration for an acoustic wave to traverse the AOM's aperture, and $\delta f$, the RF modulation bandwidth of the AOM \cite{dugan1997high}. If we denote the aperture size as $l$, then the unit effective pixel size of our AOM is given as
\begin{equation}
    \frac{l}{\delta t \delta f} = \frac{l}{(l/v_s)\delta f} = \frac{v_s}{\delta f}.
\end{equation}
With an RF bandwidth of 30 MHz and an acoustic velocity of 3.63 mm$/\mu$s, we estimate an effective unit pixel size of 120 $\mu$m, equivalent to 57 pixels across the 7 mm aperture. As the mode order increases, HG modes display finer spatial features whose sizes approach those of the AOM pixels. However, we argue that the limitation of 120 $\mu$m pixel size is not fundamental to AOM-based spatial modulation systems, and a higher mode overlap and maximal mode order are achievable using AOMs with larger clear apertures and RF bandwidths. 

Although our system is designed to work with spatial modes that are separable in the x- and y-components, the setup can access a much larger Hilbert space beyond the 24 modes shown in Fig.~\ref{fig5}(a). For instance, Laguerre-Gaussian (LG) modes can be generated by adding an HG to the LG mode converter \cite{shen2022mode}. 

\section{\label{sec:level1} Conclusion}

In conclusion, we demonstrate 2D spatial light modulation using a double AOM system. Numerical simulations were conducted to explore the impact of RF frequency adjustments on the generated spatial modes and to guide the experimental optimization process. Specifically, we demonstrate high-speed generation of 24 distinct spatial modes, each consisting of a separable two-dimensional amplitude and phase pattern defined by independent modulations along the x and y directions. The modulation rate with suitable electronics would be approximately 500 kHz, considering that the transit time of sound across the 7 mm aperture is about 2 microseconds. It is important to note that while we focus on these 24 modes, the total Hilbert space accessible by our system is significantly larger. AOM-based spatial light modulation systems of a similar kind can potentially offer pixel refresh rates of nearly 20 MHz, as reported in earlier work \cite{liu2023using}. By optimizing the RF waveform, we achieve an average mode overlap of 81\%. Using AOMs with optimized RF bandwidth and larger apertures can improve the resolution of the AOM-based spatial light modulation system and further boost the mode fidelity.

Our approach offers a promising alternative to traditional DMDs and LCoS-SLMs, which are prone to damage when modulating intense, short optical pulses. Rapid data acquisition is critical for achieving high time resolution in dynamic imaging. Thus, our system can potentially facilitate various dynamic imaging techniques, including in vivo speckle imaging and random illumination microscopy \cite{parthasarathy2008robust, boas2010laser}.

\begin{acknowledgments}

R.W.B. thanks the US Office of Naval Research, the US National Science Foundation under Award 2138174, the National Research Council Canada New Beginning Initiative Ideation Fund, and the US Department of Energy under award FWP 76295. B.B. acknowledges the support of the Banting postdoctoral fellowship of NSERC.

\end{acknowledgments}

\nocite{*}

\bibliography{reference}

\end{document}